\documentclass{aa}

\usepackage{graphics}

\begin{document}

\thesaurus{06     
          (08.16.5;  
           08.03.4;  
           02.12.3;  
           08.01.2;  
           08.13.1;  
           08.09.2 SU~Aur) 
	   }
\title{Magnetospheric Accretion and Winds on the T Tauri Star SU~Aurigae:}

\subtitle{Multi-Spectral Line Variability and Cross-Correlation Analysis.\thanks{based on observations obtained during 
the MUSICOS~96 campaign in which were involved: Isaac Newton Telescope (INT, La Palma), Observatoire de Haute Provence (OHP, France), McDonald Observatory (MDO, USA), Beijing Astronomical Observatory (BAO, Xinglong, China) and Canada-France-Hawaii Telescope (CFHT, Hawaii).}}

\author{J.M.~Oliveira \inst{1,2}
\and    B.H.~Foing \inst{1,3}
\and    J.Th.~van~Loon \inst{4}
\and    Y.C.~Unruh \inst{5}
}

\offprints{J.M.~Oliveira, ESTEC address \\{\it Correspondence to}:
joana@so.estec.esa.nl}

\institute{ESA Space Science Department, ESTEC/SCI-SO, P.O.~Box 299, NL-2200
           AG Noordwijk, The Netherlands\\
           e-mail: joana@so.estec.esa.nl, bfoing@estec.esa.nl
\and       Centro de Astrof\'{\i}sica da Universidade do Porto, Rua das
           Estrelas s/n, PT-4150 Porto, Portugal
\and       Institut d'Astrophysique Spatiale, CNRS/Univ.~Paris~XI,
           B\^{a}t.~121, Campus d'Orsay, F-91405 Orsay Cedex, France
\and       Institute of Astronomy, Madingley Road, Cambridge CB3 0HA, United
           Kingdom\\
           e-mail:  jacco@ast.cam.ac.uk	   
\and       Institut f\"{u}r Astronomie, Universit\"{a}t Wien,
           T\"urkenschanzstr. 17, A-1180 Vienna, Austria\\
           e-mail: ycu@astro.univie.ac.at
          }

\date{Received 12 May 2000 / Accepted 21 August 2000}

\titlerunning{Magnetospheric Accretion and Winds on SU~Aur}

\authorrunning{Oliveira et al.}

\maketitle

\begin{abstract}

SU~Aurigae is a T Tauri star that was well monitored during the MUSICOS~96 multi-site campaign. We present the results of the spectroscopic analysis of the  circumstellar environment of this star, particularly of the H$\alpha$, H$\beta$, \ion{Na}{i}~D and \ion{He}{i}~D3 line profiles. The signatures of modulated outflows and mass accretion events are analysed, as well as transient spectral features. We compute the cross-correlation function ($CCF$) of several pairs of (velocity bins in) spectral lines to better investigate the profiles' temporal variability. We found increasing time lags between the variability of \ion{He}{i}~D3, \ion{Na}{i}~D and H$\beta$. We propose this may be understood in terms of azimuthal distortion of the magnetic field lines due to the different rotation rates of the star and the disk. We find the slightly blueshifted absorption features in H$\alpha$, H$\beta$\ and \ion{Na}{i}~D to be anti-correlated with the accretion flow signatures. We propose that the transient absorption features in the blue wings of H$\alpha$, H$\beta$\ and \ion{Na}{i}~D (signatures of mass outflows), and flare brightenings are related to the disruption of distorted magnetospheric field lines.

\keywords{Stars: pre-main sequence --
          Circumstellar matter --
          Line: profiles --
          Stars: activity --
          Stars: magnetic fields --
          Stars: individual: SU~Aur
          }
\end{abstract}


\section{Introduction}

\subsection{Classical T Tauri Stars}

\indent T~Tauri stars (TTS) are low-mass (M~$<$~3~M$_{\sun}$) pre-main sequence stars which are normally younger than 3~$\times$~10$^{6}$ years (Appenzeller \& Mundt \cite{appenzeller}). They are of great interest as prototypes of young solar-type stars. Historically two sub-groups of TTS were defined, based on the strength of the H$\alpha$\ spectral line emission: classical T~Tauri stars (CTTS) with strong H$\alpha$\ emission and weak (or ``naked'') T~Tauri stars (WTTS). The enhanced activity of most CTTS as seen from the Balmer and metallic (permitted and forbidden) line emission is believed to be due to the presence of an accretion disk, whose UV and IR excess emission causes also broad energy distributions.\\
\indent Since the first studies of their line profiles it has been known that CTTS spectra show evidence of outflows (e.g. Herbig \cite{herbigb}). More recently also abundant observational evidence  of funnel accretion flows onto CTTS has been published (e.g. Edwards et al. \cite{edwards}).\\
\indent K\"{o}nigl (\cite{konigl})  proposed the application of the Ghosh \& Lamb (\cite{ghosh}) magnetospheric model, developed for accreting neutron stars, to explain the interaction between CTTS and their accretion disk. The accretion disk is disrupted at a few stellar radii from the central star (close to the co-rotation radius) by the presence of a stellar dipole magnetic field. According to Shu et al. (\cite{shu}) and Paatz \& Camenzind (\cite{paatz}), at the truncation point, the ionized disk material is loaded either onto inner closed magnetic field lines, accreting thereby onto the stellar surface, or it is loaded onto outer open magnetic field lines that can drive a disk-wind flow.\\
\indent  Recently, some observational support for the magnetospheric models was presented by Johns-Krull et al. (\cite{johnse}) and Guenther et al. (\cite{guenther}) who measured  kilogauss magnetic fields at the surface of CTTS, enough to effectively disrupt the accretion disk at a few stellar radii. The assumption of a stellar dipole magnetic field aligned with the rotation axis completely determines the geometry and dynamics of the disk-star interaction. According to the magnetospheric models mentioned previously, the magnetosphere is axi-symmetric with respect to the rotational axis so no rotational modulation is expected in the wind and accretion signatures, unless the disk is inhomogeneous at the truncation radius. Furthermore, variations in the inflow and outflow signatures should be approximately in phase. But a stellar magnetic field that has solely a dipolar component is unlikely and this type of models has been criticised (Safier \cite{safier}; Alencar \& Basri \cite{alencar}). The evidence for magnetospheric accretion exists (hot spots and redshifted absorption features), but theoretical models are still far from explaining the variety of observations.

\subsection{SU~Aurigae}

\indent \object{SU~Aurigae} has been classified as a CTTS (e.g. Bouvier et al. \cite{bouvier}), but also as the prototype of a separate class of T Tauri stars, the SU~Aur type stars (Herbig \& Bell \cite{herbigc}). SU~Aur has a G2 spectral type (Herbig \cite{herbiga}), thus, due to the contrast with a hotter photosphere, only H$\alpha$, and occasionally H$\beta$, appear in emission in the spectrum. \object{SU~Aur} has an exceptionally high projected rotational velocity of 60~km~s$^{-1}$\ (Johns-Krull \cite{johnsc}; Unruh et al. \cite{unruh}) and a rotational period of about 3~days. Photometric period measurements for SU~Aur are very difficult to make. For instance, Bouvier et al. (\cite{bouvier}) found multiple periods, with 2.78~days the most likely period. So far the best period estimates come from the analysis of the spectral variations of the Balmer lines. Giampapa et al. (\cite{giampapa}) found a period of 2.98~$\pm$~0.4~days in the blue wing of the H$\alpha$\ line, later confirmed also in the blue and red wings of H$\beta$\ by Johns \& Basri (\cite{johnsb}). Petrov et al. (\cite{petrov}) reported a period of 3.031~$\pm$~0.003~days measured on the red wings of the Balmer lines. Our data set suggests a shorter rotational period of about 2.6 to 2.8~days (Unruh et al. \cite{unruh}).

\subsection{Diagnostics and Models of Magnetospheric Accretion}

\indent From period analysis and equivalent width measurements of fitted components in H$\alpha$\ and H$\beta$, Johns \& Basri (\cite{johnsb}) observed that the H$\beta$\ redshifted absorption feature and the H$\alpha$\ blueshifted absorption feature (respectively accretion and wind signatures) show a periodicity of about 3~days and are approximately $180\degr$ out of phase in SU~Aur. This led them to propose a generalization of the magnetospheric model, in which the rotational axis and the magnetic dipole axis are (slightly) misaligned: the misaligned ``egg-beater'' or oblique rotator model. It predicts that the signatures of mass accretion and disk winds should be rotationally modulated and approximately in anti-phase, as they observed in SU~Aur.\\
\indent Using H$\alpha$, H$\beta$, \ion{Na}{i}~D and \ion{He}{i}~D3, we try to disentangle the two main contributions present in the line profiles, namely accretion and wind signatures. These lines probe different parts of the circumstellar environment. Thus, the temporal relation between the variability in the different lines depends on the geometry and physical conditions of the circumstellar environment, in particular of the inner disk and magnetosphere. For instance, variations in the accretion rate at the inner disk or outer magnetosphere will first affect low excitation energy features arising in this region. The higher energy lines that form in the inner accretion stream or in the accretion shock will be affected by this ``perturbation'' after a time lag related to the infall time through the magnetosphere. This delay is expected to be of the order of several hours (Smith et al. \cite{smitha}).\\
\indent Previous SU~Aur data sets typically had one to two spectra a night, even though over long time spans. Our data set, with finer time sampling, allows us for the first time to study the variations of the different components in spectral lines on short timescales. In the context of the magnetospheric model and using cross-correlation techniques, we have searched for time lags between the variability of the mentioned spectral lines. In Sect.\ 3 we describe the implementation and results of the cross-correlation analysis of our data set. In Sect.\ 4 we present our interpretations in the frame of accretion and wind processes. In Sect.\ 5 we describe two transient absorption features observed in H$\alpha$, H$\beta$\ and \ion{Na}{i}~D and we analyse their velocity evolution.


\section{Description of the SU~Aur Data Set}

\subsection{MUSICOS~96 Campaign}

\indent The MUSICOS (MUlti-SIte COntinuous Spectroscopy) network was created to provide continuous spectroscopic coverage for stellar objects, whose variability timescales demand simultaneous coverage from several longitudes (Catala \& Foing \cite{catalaa}; Catala et al. \cite{catalab}; Catala \& Foing \cite{catalac}). To fulfill this aim, existing observational capabilities were used or dedicated instruments were developed. A MUSICOS campaign is organized every two years and several stellar targets are selected for each campaign.\\
\indent Our data set was obtained in November~1996 during the MUSICOS~96 multi-wavelength campaign, that involved five different observatories: Isaac Newton Telescope (INT, La Palma), Observatoire de Haute Provence (OHP, France), McDonald Observatory (MDO, USA), Beijing Astronomical Observatory (BAO, Xinglong, China) and Canada-France-Hawaii Telescope (CFHT, Hawaii). We obtained 126 echelle spectra of SU~Aur, spanning 10.5~days, with 3.5~days covered almost continuously. H$\alpha$, \ion{Na}{i}~D and \ion{He}{i}~D3 were observed at all five observatories while H$\beta$\ could only be observed from the INT, OHP and CFHT. Thus, for this line, we obtained only 72 spectra. The data from the different observatories were rebinned to a uniform resolution of 30\ 000, or approximately 10~km~s$^{-1}$. The journals of the observations and more details about this campaign and data reduction can be found in Unruh et al. (\cite{unruh}).

\subsection{Brief Summary of Variability in the Data Set}

\begin{figure}[ht]
\resizebox{\hsize}{!}{\includegraphics{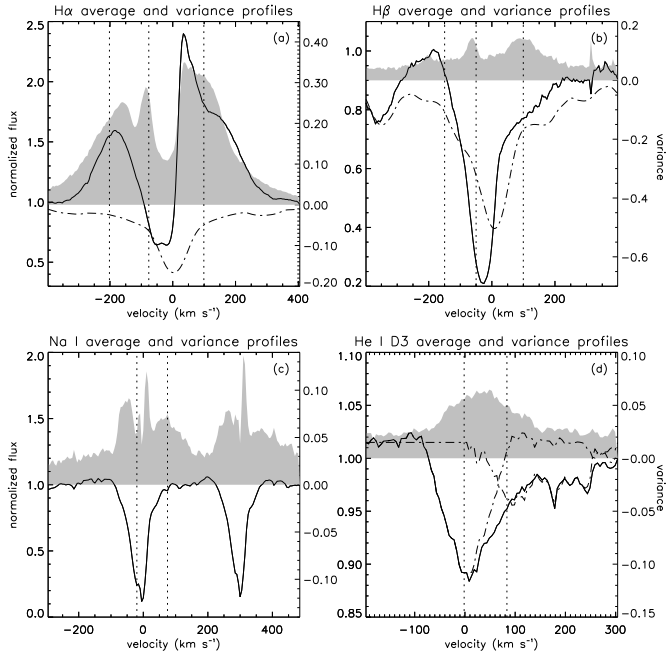}}
\caption{The average (line plots) and variance (shaded plots) of (a) H$\alpha$, (b) H$\beta$, (c) \ion{Na}{i}~D and (d) \ion{He}{i}~D3. The y-axis scale on the left is for the average and on the right for the variance. For H$\alpha$\ and H$\beta$\ a rotationally broadened template profile is plotted (dot-dashed line in (a) and (b)) as a photospheric reference. The narrow feature that appears in the variance profiles of H$\alpha$, H$\beta$\ and \ion{Na}{i}~D, between $-150$ and $-50$~km~s$^{-1}$, is the signature of transient outflows (Sect.\ 5). The sharply peaked central feature visible in the variance profile of \ion{Na}{i}~D is related to the narrow interstellar component. The two components present in the \ion{He}{i}~D3 profile are also plotted (dot-dashed line in (d)); the smaller features appearing at 180 and 230~km~s$^{-1}$ are telluric water residues. The vertical dotted lines appearing in all the graphs are the velocity bins chosen for the $CCF$ analysis (Sect.\ 3.2); the binsize is 5~km~s$^{-1}$.}
\label{f1}
\end{figure}

\indent We highlight in this subsection the main characteristics of the variability in our data set. The period analysis was performed by Unruh et al. (\cite{unruh98}, \cite{unruh}). They computed the periodograms of the above mentioned spectral lines (see figures in Unruh et al. \cite{unruh}). We independently determined the following periods: 2.63~days in \ion{He}{i}~D3 [$-$75:175]~km~s$^{-1}$ and in the red wings of \ion{Na}{i}~D [30:150]~km~s$^{-1}$, and 2.80~days in the red wing of H$\beta$ [50:175]~km~s$^{-1}$ and in the far blue wing of H$\alpha$ [$-$375:$-$175]~km~s$^{-1}$. This might indicate that the regions of formation of these lines are not strictly in co-rotation. However, the peaks in the periodograms are fairly broad, thus these periods are all consistent with a single periodicity, between 2.5 and 3~days.\\
\indent In the periodogram of H$\alpha$, and less conspicuously in the periodograms of H$\beta$\ and \ion{Na}{i}~D, there is a broad peak at about 5~days that we believe to be related to the flux enhancement observed in H$\alpha$ and H$\beta$ at roughly the middle of the time series (Sect.\ 3.2). We have computed the periodograms of the spectral lines, removing the spectra affected by this flux enhancement. We confirm our period determinations and, in fact, the peak at 5~days vanishes.\\
\indent In Fig.~\ref{f1} we show the average and variance profiles of H$\alpha$, H$\beta$, \ion{Na}{i}~D and \ion{He}{i}~D3. We also plot a rotationally broadened photospheric template for the two Balmer lines to indicate clearly the circumstellar contribution (Fig.~\ref{f1}a,b). In both Balmer lines and in \ion{Na}{i}~D transient outflows are present. These appear in the variance profiles of these lines as the peaks between $-150$ and $-50$~km~s$^{-1}$. Small mismatches in the wavelength calibration between the different sites and the rebinning of the data cause the narrow features in the \ion{Na}{i}~D variance profile at the position of the interstellar absorption lines.\\
\indent The average profile of \ion{He}{i}~D3 is very asymmetric and appears to be formed by two components, one at rest velocity and a redshifted component (Fig.~\ref{f1}d). The rest component is obtained, in a crude decomposition, by reflecting the blue wing of the average profile about zero velocity. The redshifted component is isolated by subtracting this reflection from the average profile. This component is centered at approximately 83~km~s$^{-1}$. The two components are broader than 2~$\times$~v $\sin$ i and the redshifted component extends at least until 200~km~s$^{-1}$.


\section{Multi-Line Cross-Correlation Analysis}

\begin{figure*}[ht]
\resizebox{\hsize}{17.5cm}{\includegraphics{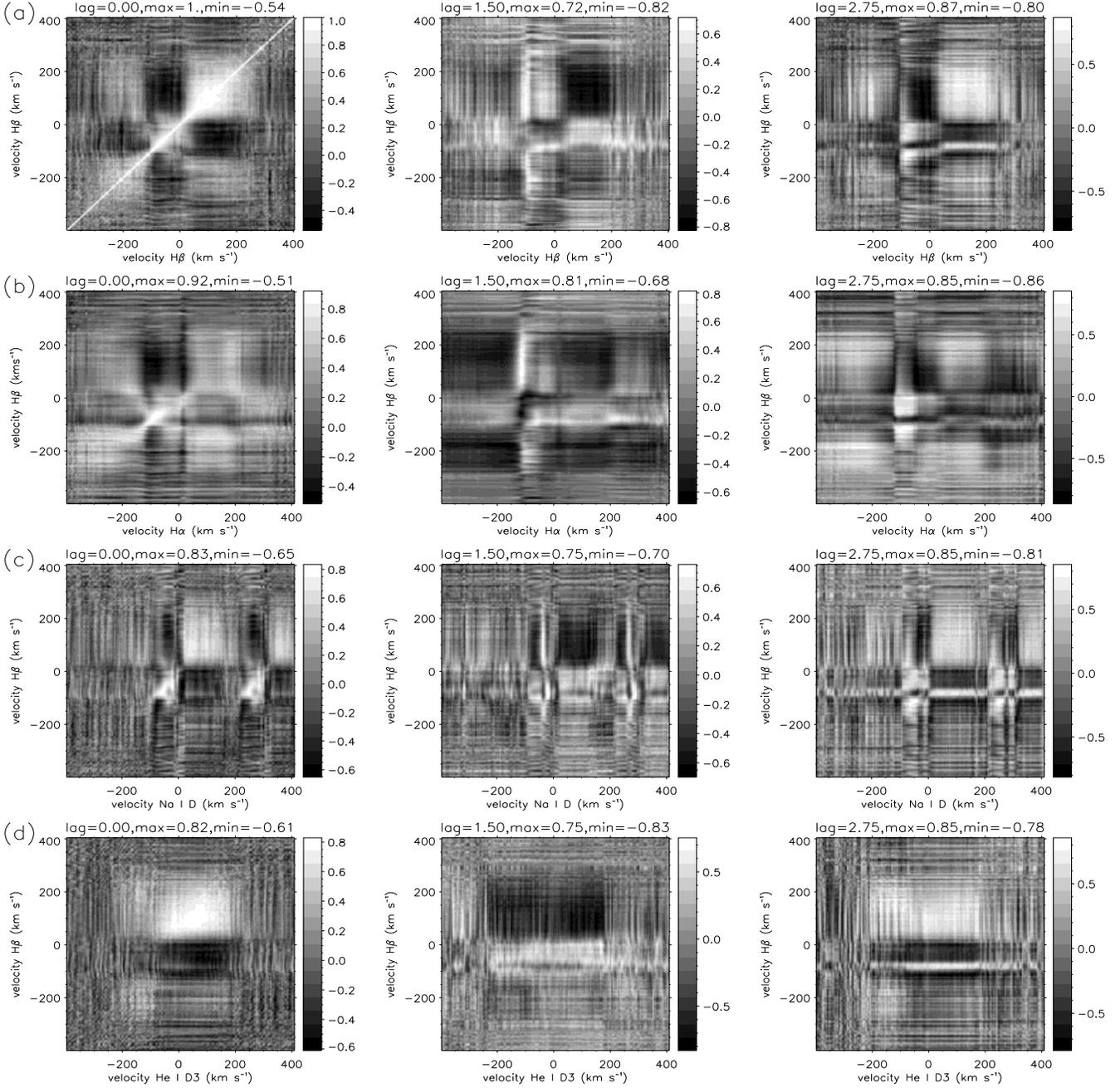}}
\caption{Cross-correlation function intensity maps: (a) $ACF$(H$\beta$), (b) $CCF$(H$\alpha$,H$\beta$), (c) $CCF$(\ion{Na}{i},H$\beta$) and (d) $CCF$(\ion{He}{i},H$\beta$). Each row represents intensity maps for one pair of lines, for 3 time lags: 0, $\sim$~P/2 and $\sim$~P. The period used is approximately 2.75~days. Bright represents high correlation coefficients and dark strong anti-correlations. The intensity (grey) scale is not the same in all maps. In each map, the time lag ($\Delta \tau$) is indicated as well as the maximum/minimum correlation coefficients found.}
\label{f2}
\end{figure*}

\indent The spectrum of SU~Aur is rich in spectral lines thought to originate from the circumstellar environment of the star. The Balmer lines, H$\alpha$\ and H$\beta$, the \ion{Na}{i}~D and \ion{He}{i}~D3 lines, due to their different physical properties, probe different parts of the circumstellar material. Furthermore, in these lines mingle several different components: Giampapa et al. (\cite{giampapa}) and Johns \& Basri (\cite{johnsb}) proposed that the profiles of, respectively, H$\alpha$\ and H$\beta$\ could be decomposed in the contributions of an enhanced chromosphere, and a slow and a fast wind components.\\
\indent The total time span of our data set allows us to sample variability related to rotational modulation. With our fine time sampling, we also expect to characterize variability on much shorter time scales, for instance arising from the inner magnetosphere. We intend to analyse how variability propagates through the different spectral lines. In our analysis of these complex profiles, we chose the cross-correlation function that expresses how the correlation coefficient between two time series varies with the time delay between them.\\
\indent Cross-correlation techniques have been widely used in variability studies of active galactic nuclei (White \& Peterson \cite{white}). Smith et al. (\cite{smithb}) applied this method to the equivalent width of the Balmer and \ion{Ca}{ii} H~\&~K lines for several T~Tauri stars, including SU~Aur, but their low-resolution spectra did not allow to disentangle the different components that mingle in those spectral lines. This technique was also previously used, for instance, in Johns \& Basri (\cite{johnsa}) for the analysis of H$\alpha$\ in several CTTS (also SU~Aur) and Johns-Krull \& Basri (\cite{johnsd}) for the analysis of several lines in another TTS \object{DF~Tau}. \\
\indent We calculated the cross-correlation function ($CCF$) of pairs of spectral lines as a function of the time lag $\tau$ as well as the auto-correlation functions ($ACF$). In Appendix~A we define the $CCF$ and describe its implementation for our type of time series, as well as the error analysis.

\subsection{Application of $CCF$ Analysis to SU~Aur Data}

\indent Using the technique described in Appendix~A, we computed for each pair of spectral lines a cross-correlation function intensity map that allows a first glimpse at the time lag information contained in the time series. These maps were calculated for each pair of spectral lines for lags multiple of 0.25~days (6~h). In Fig.~\ref{f2} we present the most striking examples of the results we obtained. Each column in this figure represents the $CCF$ computed at three representative time lags: 0, $\sim$~P/2 and $\sim$~P, where P is the rotational period, taken to be approximately 2.75~days. It can be seen that the periodicity in the spectral lines is reflected in the behaviour of the $CCF$s with time lag. The $CCF$s vary from high positive correlation coefficient to high negative correlation coefficient in a periodic way (e.g. Fig.~\ref{f2}a). These maps illustrate the time relation between the different spectral lines (Oliveira et al. \cite{oliveirab}). The red wing of H$\beta$\ seems to vary approximately in phase with the red wings of H$\alpha$\ and \ion{Na}{i}~D (Fig.~\ref{f2}b,c respectively), and with the whole \ion{He}{i}~D3 profile (Fig.~\ref{f2}d). Furthermore, the blue and red wings of H$\beta$\ seem out of phase (Fig.~\ref{f2}a).\\
\indent These kind of maps cannot be used to determine time lags with enough precision. Their computation is too time consuming to be a viable tool for a finer analysis with better time sampling. Therefore we have decided to choose and isolate velocity bins in the spectral lines that are representative of the profiles' variations.

\subsection{Velocity Bins for the CCF}

\begin{figure}[ht]
\resizebox{\hsize}{12cm}{\includegraphics{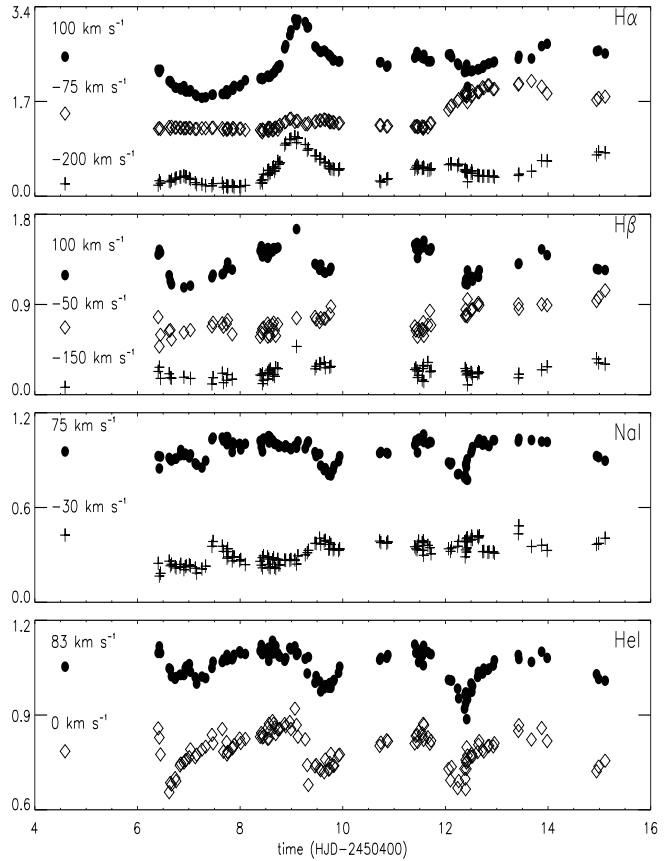}}
\caption{Intensity variations of the chosen velocity bins. From top to bottom: the variations of H$\alpha$, H$\beta$, \ion{Na}{i}~D1 and \ion{He}{i}~D3, as indicated on the upper corner of each graph. Also indicated is the corresponding velocity, on the left side. The flux enhancement is clearly visible in the red and far blue wings of both Balmer lines. In the slightly blueshifted bin of both Balmer lines, from t~$\sim$~12 onwards, the effect of the second transient feature can be seen (Sect.\ 5).}
\label{f3}
\end{figure}

\begin{figure}[h]
\resizebox{\hsize}{10cm}{\includegraphics{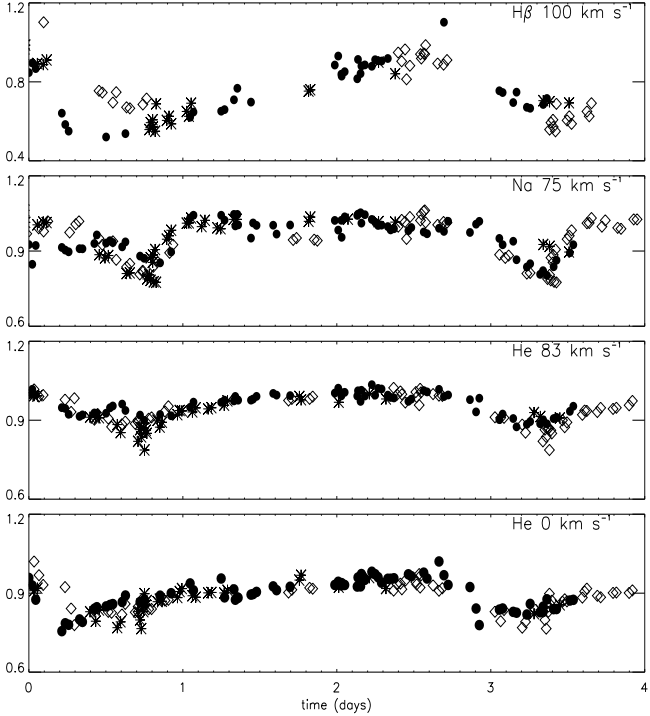}}
\caption{Intensity variations folded in time. From top to bottom: H$\beta$\ at 100~km~s$^{-1}$, \ion{Na}{i}~D1 at 75~km~s$^{-1}$, \ion{He}{i}~D3 at 83 and 0~km~s$^{-1}$. The intensity variations are plotted against: $\bullet$\ time, $\diamond$\ (time~$-$~P), $\ast$\ (time~$-$~2$\times$P). The period used was P~=~2.63~days. Obviously, the variability of H$\beta$\ has other non-periodic components. Folded in this way, the flux variations in \ion{He}{i}~D and \ion{Na}{i}~D clearly show their periodicity. The lagged behaviour of the variations of the three spectral lines is evident.}
\label{f4}
\end{figure}

\indent We selected coherent velocity domains by analysing the $ACF$ for each spectral line. This is very well illustrated in Fig.~\ref{f2}a for H$\beta$. In this $ACF$, we can see that the H$\beta$\ profile is composed of three distinct domains: \mbox{[$-$300:$-$125]}~km~s$^{-1}$, \mbox{[$-$125:25]}~km~s$^{-1}$ and \mbox{[25:250]}~km~s$^{-1}$. Within these velocity domains the profile behaves in a concerted way. In each such interval we selected a representative velocity bin for more detailed $CCF$ analysis: $-$150~km~s$^{-1}$ (the far blue wing), $-$50~km~s$^{-1}$ (the slightly blueshifted absorption component) and 100~km~s$^{-1}$ (the redshifted absorption component). For H$\alpha$\ three domains are also defined: \mbox{[$-$400:$-$125]}~km~s$^{-1}$, \mbox{[$-$125:25]}~km~s$^{-1}$\ and \mbox{[25:250]}~km~s$^{-1}$; the appropriate bins are then $-$200~km~s$^{-1}$, $-$75~km~s$^{-1}$ and 100~km~s$^{-1}$, corresponding to the same regions in H$\beta$. For \ion{Na}{i}~D the blue and red wings seem to vary independently; we have thus chosen two velocity bins: $-$30~km~s$^{-1}$\ and 75~km~s$^{-1}$. The profile of \ion{He}{i}~D3 varies as a whole within the interval [$-$75:150]~km~s$^{-1}$. Still, the \ion{He}{i}~D3  profile seems composed of two components (Sect.\ 2.2). Thus, we will investigate their behaviour separately, selecting bins at 0 and 83~km~s$^{-1}$. The selected bins for these spectral lines are shown in Fig.~\ref{f1}. In Fig.~\ref{f3} we plot the intensity variations of the velocity bins that we have selected for each line: from top to bottom, H$\alpha$, H$\beta$, \ion{Na}{i}~D1 and \ion{He}{i}~D3.\\
\indent The H$\alpha$\ intensity variations clearly show the flux enhancement seen both in the blue and red wings of the profile. The enhancement lasted for a minimum of 2~days, and the peak intensity occurs at about t~=~9~days. The increase was $\sim 65$ \% in the blue wing and $\sim 72$ \% in the red wing. The central absorption feature ($-75$~km~s$^{-1}$) shows, from t~$\sim$~12 onwards, the presence of the second transient features (see Sect.\ 5). In the H$\alpha$\ flux variations there are hardly any visible traces of the $\sim$~3~day periodicity.\\
\indent The analysis of the flux enhancement in  H$\beta$\ is more complicated as there are fewer data points. Still, the flux enhancement can also be seen, causing variations of $\sim 22$~\% in the far blue wing and $\sim 37$~\% in the red wing. In the red wing of H$\beta$, the $\sim$~3~day periodicity can be seen and in the slightly blueshifted  absorption feature there are some indications of this too. Furthermore, the quasi anti-phase behaviour of these two parts of the H$\beta$\ profile can be observed. The variations of the far blue wing of H$\beta$, with the exception of the flux enhancement, show no relation with the variations in the rest of the profile.\\
\indent Both components of \ion{He}{i}~D3 and the red wing of \ion{Na}{i}~D1, show clear 2.63~days periodicity. The variations are not sinusoidal but the periodic deepening of the absorption features is clear. The behaviour of the blue and red wings of \ion{Na}{i}~D1 seems anti-correlated.\\
\indent In order to clarify and compare more easily the variability of the four velocity bins where signatures of periodicity were seen, we folded in time the intensity variations of H$\beta$\ at 100~km~s$^{-1}$, \ion{Na}{i}~D1 at 75~km~s$^{-1}$ and \ion{He}{i}~D3 at 0 and 83~km~s$^{-1}$. This is done in Fig.~\ref{f4} by plotting the intensity variations as a function of time, (time~$-$~P) and (time~$-$~2$\times$P), starting at t$_{0}$~=~6.4~days. The period used was the one determined for \ion{He}{i}~D and \ion{Na}{i}~D: P~=~2.63~days. We have not done this for any bin in H$\alpha$\ because large sporadic changes seem to dominate. In H$\beta$\ there is also some non-periodic variability. The behaviour of the other two lines is mainly controlled by the phenomenon with periodicity of $\sim$~2.63~days. Very interesting in Fig.~\ref{f4} is that it makes clearly visible the time-lagged behaviour between the variability of \ion{He}{i}~D3 at 0 and 83~km~s$^{-1}$, \ion{Na}{i}~D1 at 75~km~s$^{-1}$ and H$\beta$\ at 100~km~s$^{-1}$. In the remainder of this section we calculate these time lags and relate them to accretion and wind flows.

\subsection{Time Lags between Multi-Line Variations}

\begin{figure*}[ht]
\resizebox{\hsize}{18cm}{\includegraphics{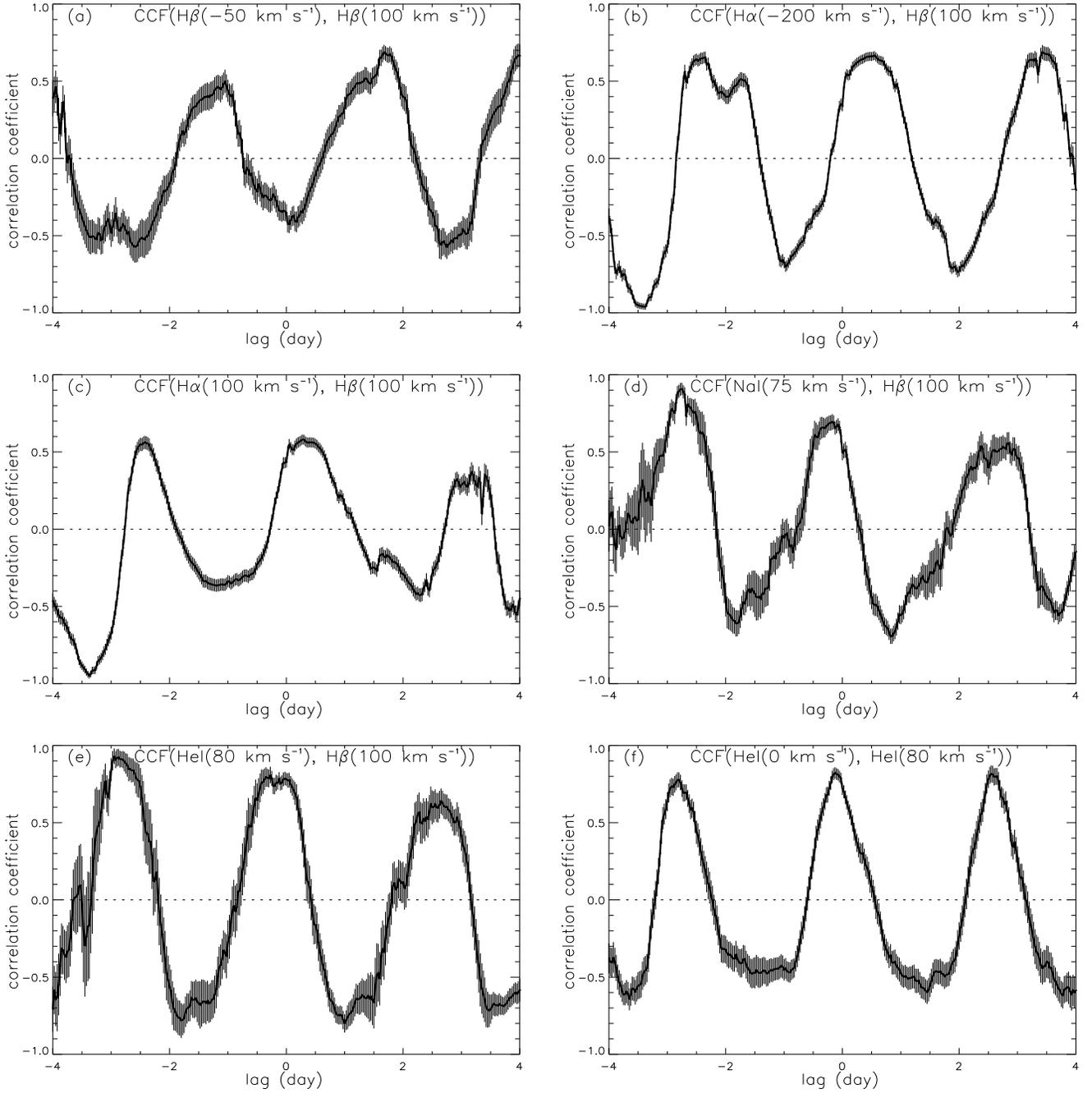}}
\caption{Cross-correlation as a function of time lag. In each graph the $CCF$  of a pair of selected velocity bins is plotted with the error bars obtained from the Monte Carlo simulations; the spectral lines and the velocity bins used are indicated. The $CCF$s were calculated with a sampling of $\Delta \tau=0.025$~day. The recurrence of the peaks is a clear result of the periodicity present in the spectral lines. The amplitude and shape of the $CCF$ reflect how strong was the periodicity detected in the velocity bins.}
\label{f5}
\end{figure*}

\begin{table*}[ht]
\begin{center}
\begin{tabular}{lc|lc|cc|ccc|l}
\hline
line~1                         &
v~1                            &
line~2                         &
v~2                            &
peak                           &
amplitude                      &
median ($\tau$)                &
$\tau -\Delta \tau _{67}$      &
$\tau +\Delta \tau _{67}$      &
FAP                            \\
                               &
(km s$^{-1}$)                  &
                               &
(km s$^{-1}$)                  &
(days)                         &
                               &
(days)                         &
(days)                         &
(days)                         &
peak                           \\
\hline
H$\beta$                       &
   \llap{$-$5}0                &
H$\beta$                       &
   \llap{10}0                  &
          1.675                &
	  0.62                 &
          1.670                &
          1.654                &
          1.762                &
          $2 \times 10^{-5}$   \\
H$\alpha$                      &
  \llap{$-$20}0		       &
H$\beta$		       &
   \llap{10}0                  &
          0.550                &
	  0.63		       &
	  0.525		       &
	  0.422		       &
	  0.544		       &
          $4 \times 10^{-7}$   \\	  
H$\alpha$                      &
  \llap{10}0		       &
H$\beta$		       &
   \llap{10}0                  &
          0.275                &
	  0.55		       &
	  0.272		       &
	  0.192		       &
	  0.406		       &
          $5 \times 10^{-5}$   \\
\ion{Na}{i}	               &
   \llap{7}5                   &	  
H$\beta$		       &
   \llap{10}0                  &	  
   \llap{$-$}0.175	       &  
             0.64	       &
   \llap{$-$}0.188	       &
   \llap{$-$}0.288	       &
   \llap{$-$}0.095	       &
   	  $3 \times 10^{-7}$   \\
\ion{He}{i}		       &
   \llap{8}0                   &
H$\beta$		       &
   \llap{10}0                  &
   \llap{$-$}0.275             &
             0.72	       &
   \llap{$-$}0.273	       &
   \llap{$-$}0.356	       &
   \llap{$-$}0.027	       &
          $3 \times 10^{-9}$   \\	   	  
\ion{He}{i}                    &
           0		       &
\ion{He}{i}		       &
   \llap{8}0                   &
   \llap{$-$}0.125             &
             0.69	       &
   \llap{$-$}0.130	       &
   \llap{$-$}0.162	       &
   \llap{$-$}0.071	       &
          $2 \times 10^{-16}$  \\
\ion{He}{i}		       &
   \llap{8}0                   &
H$\beta$		       &
   \llap{$-$15}0               &
   \llap{$-$}0.650             &
             0.47              &
   \llap{$-$}0.725	       &
   \llap{$-$}0.967             &
   \llap{$-$}0.580	       &
          $3 \times 10^{-3}$   \\		  
\ion{He}{i}		       &
   \llap{8}0                   &
H$\beta$		       &
   \llap{$-$5}0                &
   \llap{$-$}1.750             &
             0.51	       &
   \llap{$-$}1.770	       &
   \llap{$-$}1.843	       &
   \llap{$-$}1.341	       &
          $1 \times 10^{-3}$   \\		  
\ion{He}{i}		       &
   \llap{8}0                   &
H$\alpha$		       &
   \llap{$-$20}0               &
   \llap{$-$}0.650             &
             0.52	       &
   \llap{$-$}0.710	       &
   \llap{$-$}0.861	       &
   \llap{$-$}0.619	       &
          $2 \times 10^{-7}$   \\
\ion{He}{i}		       &
   \llap{8}0                   &
H$\alpha$		       &
   \llap{$-$7}5                &
   \llap{$-$}1.900             &
             0.38	       &
   \llap{$-$}1.906	       &
   \llap{$-$}1.946	       &
   \llap{$-$}1.803	       &
          $4 \times 10^{-3}$   \\	  
\ion{He}{i}		       &
   \llap{8}0                   &
H$\alpha$		       &
   \llap{10}0                  &
   \llap{$-$}0.650             &
             0.34	       &
   \llap{$-$}0.620	       &
   \llap{$-$}0.670	       &
   \llap{$-$}0.516	       &
          $4 \times 10^{-3}$   \\
\ion{He}{i}		       &
           0                   &
\ion{Na}{i}		       &
   \llap{7}5                   &
   \llap{$-$}0.075             &
             0.65	       &
   \llap{$-$}0.079	       &
   \llap{$-$}0.097	       &
   \llap{$-$}0.058	       &
          $4 \times 10^{-14}$  \\
\ion{He}{i}		       &
   \llap{8}0                   &
\ion{Na}{i}		       &
   \llap{$-$3}0                &
   \llap{$-$}2.050             &
             0.40              &
   \llap{$-$}1.883             &
   \llap{$-$}2.097             &
   \llap{$-$}1.722             &
          $2 \times 10^{-3}$   \\	  
\ion{He}{i}		       &
   \llap{8}0                   &
\ion{Na}{i}		       &
   \llap{7}5                   &
             0.050             &
             0.72	       &
             0.038	       &
             0.026	       &
             0.058	       &
          $1 \times 10^{-18}$  \\
H$\alpha$                      &
  \llap{$-$7}5		       &
H$\beta$		       &
   \llap{$-$5}0                &
   \llap{$-$}0.125             &
             0.76	       &
   \llap{$-$}0.130	       &
   \llap{$-$}0.254	       &
   \llap{$-$}0.047	       &
          $1 \times 10^{-11}$  \\	
H$\alpha$		       &
   \llap{$-$20}0               &
H$\alpha$		       &
   \llap{10}0                  &
   \llap{$-$}0.075             &
             0.88	       &
   \llap{$-$}0.087	       &
   \llap{$-$}0.096	       &
   \llap{$-$}0.079	       &
          $1 \times 10^{-18}$  \\
H$\alpha$		       &
   \llap{$-$20}0               &
H$\beta$		       &
   \llap{$-$15}0               &
   \llap{$-$}0.075             &
             0.76	       &
   \llap{$-$}0.084	       &
   \llap{$-$}0.110             &
   \llap{$-$}0.052	       &
          $1 \times 10^{-11}$  \\	  
\hline
\end{tabular}
\end{center}
\caption{Time lags found for several pairs of velocity bins. Pairs
that do not yield significant lags or provide redundant information are not listed. $CCF$(l$_{1}$,l$_{2}$) is the cross-correlation function of l$_{1}$ and l$_{2}$. In the first four columns, the spectral lines and velocity bins are given. The peak and amplitude of the $CCF$ are listed in columns 5 and 6, the median ($\tau$) and $\tau \pm \Delta \tau _{67}$\ of the cross-correlation probability distribution in columns 7, 8 and 9. The statistical significance of the peak by means of the false alarm probability (FAP) is given in the last column. The first six $CCF$s are listed in the order as they appear in Fig.~\ref{f5}.}
\label{t1}
\end{table*}

\indent We compute the $CCF$ for each pair of chosen velocity bins with a sampling $\Delta \tau~=~0.025$~day (approximately half of the maximum exposure time) for time lags ranging from $-$4 to 4 days (for larger time lags the number of points in the calculation would be too small). For each lag $\tau$\ the $CCF$ is computed with a different number of pairs of points, so the statistical significance of a given value of the correlation coefficient varies with the lag. In our analysis of the $CCF$s we will consider only peaks significant at more than the 99~\% level. We have computed the false alarm probability following Bevington \& Robinson (\cite{bevington}).\\
\indent In Fig.~\ref{f5} we show some of the $CCF$s for the pairs of velocity bins that give the most interesting results. In each plot, the $CCF$ is represented with error bars that are obtained from the standard deviation of the Monte Carlo simulations. The periodicity of the $CCF$s is evident and is a consequence of the periodic variations in the spectral lines analysed. Taking for example Fig.~\ref{f5}f, there is a central peak of the $CCF$ at about $\tau$~=~0, and two more peaks respectively at $\pm$~P. For velocity bins where only a weak peak was detected in the period analysis, e.g. the  blue wing of H$\beta$\ and the red wing of H$\alpha$, that is reflected in the amplitude (i.e. correlation coefficient) and shape of their $CCF$s with respect to the red wing of H$\beta$\ (Fig.~\ref{f5}a,c). On the contrary, the $CCF$s between clearly periodic time series show sharp peaks with high correlation coefficient, as it is seen for the $CCF$ between the redshifted component of \ion{He}{i}~D3 and the red wing of H$\beta$ (Fig.~\ref{f5}e).\\
\indent The shape of the $CCF$s could be interpreted as meaning that there are features which appear around a limb of the star, last coherently for half a rotation and then disappear around the other limb. The fact that the ``peaks'' and ``valleys'' have similar amplitude could indicate that features on both sides of the star behave similarly. The exception is the case of the two \ion{He}{i}~D3 components (Fig.~\ref{f5}f).\\
\indent In Table~\ref{t1} we list the relevant time lags found for each pair of velocity bins (only the peak closer to zero was analysed in detail). Pairs of velocity bins that did not yield significant lag information or were redundant were discarded. We give both the peak and amplitude of the $CCF$, the median of the cross-correlation peak distribution ($CCPD$) and the intervals $\tau \pm \Delta \tau _{67}$, obtained by interpolation of the $CCPD$ (Appendix~A). Also listed is the peak's significance by means of the false alarm probability (FAP). \\
\indent In this table there are positive and negative lags. This is a matter of convention in the way the $CCF$ is computed. In our case if $CCF$(l$_{1}$,l$_{2}$) yields a lag $\tau>0$ this means that l$_{2}$ is $\tau$ in advance of l$_{1}$; conversely if $CCF$(l$_{3}$,l$_{2}$) yields a lag $\tau<0$ it means that l$_{3}$ is $|\tau|$ in advance of l$_{2}$. This is relevant when several (velocity bins of) different lines are compared and sorted in time.\\
\indent The far blue and red wings of H$\alpha$ (respectively $-$200 and 100~km~s$^{-1}$) correlate well, with a single broad peak at approximately zero lag. Also well correlated at zero lag with these H$\alpha$\ regions is the far blue wing of H$\beta$\ ($-$150~km~s$^{-1}$). This is probably due to the contribution of a turbulent broadening mechanism (e.g. Basri \cite{basri}; Edwards et al. \cite{edwards}; Alencar \& Basri \cite{alencar}). The peak of the $CCF$ between the red wings of the two Balmer lines is at approximately 0.3~days (Fig.~\ref{f5}c), indicating a delay between the variability effect on the two lines. The $CCF$ of the blueshifted absorption feature in H$\alpha$\ ($-75$~km~s$^{-1}$), and the equivalent feature in H$\beta$\ ($-50$~km~s$^{-1}$), shows a very broad peak at approximately lag $\tau$~=~0. As a whole, the H$\alpha$\ profile correlates less well than H$\beta$\ with the other lines. This is because the different effects mingle more in H$\alpha$\ as it originates in a larger volume. The blue and red absorption features in H$\beta$\ ($-$50 and 100~km~s$^{-1}$ respectively) are anti-correlated at lag $\tau$~=~0, and show two peaks at lag $\tau$~=~$\pm$~P/2 (Fig.~\ref{f5}a).\\ 
\indent The red absorption features of \ion{Na}{i}~D correlate well with the red absorption feature in H$\beta$\ (Fig.~\ref{f5}d) and less well with the red wing of H$\alpha$. The blue absorption wings of \ion{Na}{i}~D are well correlated with the slightly blueshifted absorption features in the Balmer lines. At approximately zero lag, the blueshifted regions of these spectral lines are anti-correlated with the red wings of the same lines.\\
\indent There seems to be a small lag between the variability that affects the two \ion{He}{i}~D3 components, respectively 0 and 80~km~s$^{-1}$ (Fig.~\ref{f5}f). In agreement with this, the $CCF$s of the redshifted absorption features in \ion{Na}{i}~D and the two \ion{He}{i}~D3 components also show slightly different peaks. However, the other spectral lines, when correlated with the \ion{He}{i}~D3 components, yield peaks of the $CCF$ that are consistent with a single time lag. Therefore, in Table~\ref{t1} we have only listed the $CCF$ information on one of the \ion{He}{i}~D3 components, the component at 80~km~s$^{-1}$. We believe that the time lag detected between the two \ion{He}{i}~D3 components is real but small, thus not always detectable. Both \ion{He}{i}~D3 components are well correlated with the red wing of H$\beta$\ (Fig.~\ref{f5}e) and \ion{Na}{i}~D. The slightly blueshifted absorption features in H$\alpha$, H$\beta$\ and \ion{Na}{i}~D are anti-correlated at lag $\tau$~=~0 with \ion{He}{i}~D3 and their $CCF$ shows then two peaks at lag $\tau$~=~$\pm$~P/2 (Table~\ref{t1}).\\
\indent Between the redshifted absorption component in \ion{Na}{i}~D and the rest absorption component of \ion{He}{i}~D3 a time lag of $-$0.075~days was found, indicating the advance of the variability in this \ion{He}{i}~D3 component. On the other hand, the variability in the redshifted \ion{Na}{i}~D component seems to anticipate the variability in the redshifted absorption component in \ion{He}{i}~D3 by 0.050~days. This time lag is at the limit of our time resolution, so we believe the variability in these components to be practically simultaneous. The variability of the redshifted absorption component of H$\beta$\ is delayed by about 0.28~days with respect to the variability in the redshifted \ion{He}{i}~D3 component.\\
\indent In summary, the absorption features in the red wings of H$\beta$ and \ion{Na}{i}~D and the \ion{He}{i}~D3 line correlate strongly over a large velocity range. The correlation with the red wing of H$\alpha$\ is less clear. Anti-correlated with the red wings of H$\beta$\ and \ion{Na}{i}~D, and with \ion{He}{i}~D3, are the absorption features in the near blue wings of H$\alpha$, H$\beta$\ and \ion{Na}{i}~D. The far blue (emission) wings of both Balmer lines are well correlated with each other. We found increasing time lags between the rest absorption component of \ion{He}{i}~D3, the redshifted absorption components of \ion{Na}{i}~D and \ion{He}{i}~D3 and the redshifted H$\beta$\ absorption component. 


\section{Interpretation: Accretion, Winds and Disk}

\subsection{Multi-line Time Lags in SU~Aur}

\indent Normally, in stars of spectral type G2 there is no \ion{He}{i}~D3 line. The conditions for the formation of the \ion{He}{i}~D3 line (high temperature or density) are consistent with this line being formed at the footpoints of the accretion column or slightly above the photosphere, where the kinetic energy is released, or by XUV back-warming of the surface by an active corona. For most CTTS the \ion{He}{i}~D3 line appears in emission but in SU~Aur this line is in absorption. We decompose the \ion{He}{i}~D3 average profile in two components (Fig.~\ref{f1}d). The component centered at rest velocity probably originates in post-shock gas at the base of the accretion column, and the redshifted component in the accretion column (Najita et al. \cite{najita}). We have some indication that there is a time-lagged behaviour between the two components, and this would also be consistent with the two lines being formed in slightly displaced regions. But, the fact that we observe the same period (2.63~days) and that they vary almost in synchrony confirms that the formation of both components is closely linked.\\
\indent The absorption component in the red wings of \ion{Na}{i}~D and the \ion{He}{i}~D3 line vary in an almost synchronized way, implying that the \ion{Na}{i}~D lines are also formed close to the star. There is a small delay in the variability observed in \ion{He}{i}~D3 and the variability observed in the absorption component in the red wing of H$\beta$. Magnetospheric models (e.g. Hartmann et al. \cite{hartmanna}) strongly suggest that the redshifted absorption features in the Balmer and metallic lines in CTTS are formed in the accretion funnels. We conclude then that the \ion{Na}{i}~D and H$\beta$\ redshifted absorption features appear to arise in the inner accretion flow, with H$\beta$\ forming furthest away from the star. In H$\alpha$\ the situation is more ambiguous as there is a large bulk emission whose origin is uncertain. Still, the redshifted absorption component in H$\alpha$\ is probably also related with the accretion flow, but at larger distances from the star, and the time lags we observe are consistent with this.\\
\indent The time lags we determined suggest that a perturbation is seen in succession in the central absorption component in \ion{He}{i}~D3, in the red wing of \ion{Na}{i}~D and the redshifted component of \ion{He}{i}~D3, the red wing of H$\beta$\ and finally the red wing of H$\alpha$. We should emphasize two things. First, these absorption features do not appear at the high velocities characteristic of free-fall. This is consistent with viewing SU~Aur close to edge-on, i.e.\ under a small angle with the plane of the disk: hence the line-of-sight always makes a considerable angle with the base of the accretion flow probably at high latitudes on the star. Second, the time-lagged behaviour that we observe cannot be due to time-variable (non-steady) accretion, because in that case the lines that form closer to the disk would be affected first and that is the reverse of what we observe. In the oblique magnetospheric model, the column density of the accreting matter seen in absorption against the star varies with the viewing phase of the system: at a certain phase we look through the densest parts of the accretion flow, whilst half a rotation later the densest parts are no longer seen in front of the star (see Johns \& Basri \cite{johnsb}). In this scenario no time-lagged behaviour will be observed if the magnetosphere co-rotates with the star as a solid axi-symmetric body, due to the radial alignment of the regions of formation of the different spectral lines.\\
\indent Smith et al. (\cite{smithb}) found a similar sequence of increasing time lags for decreasing Balmer series, on one night of observations, that they attribute to an accretion stream and shock seen as they moved over the limb of the star. However, any such occultation effects will be time-symmetric: if the ingress of a region of enhanced column density is seen subsequentially in spectral lines of different excitation conditions, then the reverse sequence is seen for the egress of that region. As a result, although the duration of eclipse may differ from line to line, the time of mid-eclipse will be equal for all spectral lines. Similarly, the inner regions of the densest part of the accretion flow appear longer in front of the star than the outer regions, but the times of mid-transit are the same.\\
\indent  We propose that the time-lagged behaviour that we observe may occur if the anchors of the magnetic field lines in the star have a higher rotation rate than those in the inner disk. Then the field lines and hence the accretion flow are no longer radial and the time symmetry for ingress/egress of transit (or occultation) events is broken: mid-transit (or mid-eclipse) will occur at different times for spectral lines formed at different distances to the star, causing the observed time lags. This seems also to be supported by a possible increase of modulation period with distance from the star (see Sect.\ 2.2), suggesting that the inner parts of the accretion disk are rotating slower than the stellar surface. Thus there is an external torque working on the magnetosphere that would wind up the magnetic field lines. Apparently an equilibrium is established between the external torque and the magnetic pressure contained in the azimuthally distorted magnetic field (see below). It would also hint at a somewhat smaller stellar rotation period of P~$\sim2.63$~days, more easily reconciled with a smaller stellar radius than the usually adopted R$_\star~\sim~3.6$~R$_{\sun}$, which is also suggested by recent modelling of the spectral energy distribution (Oliveira  et al. \cite{oliveirac}).\\
\indent A time lag $\tau\sim$~1.67~day separates the variations in the absorption features in the red and blue wing of H$\beta$. A slightly bigger time lag (1.77~days) is observed between the blue absorption feature in H$\beta$\ and the \ion{He}{i}~D3 line. Both the blueshifted absorption features in H$\alpha$\ and \ion{Na}{i}~D show similar lagged behaviour (time lags of respectively 1.90 and 1.88~day) with respect to \ion{He}{i}~D3. In the scenario of the magnetospheric models, these blueshifted absorption features are wind indicators. Thus, the signatures of accretion and disk winds are out of phase. Our $CCF$ calculations indicate a phase shift of about 240$\degr$, but, as it can see from Table~\ref{t1}, the peaks are broad and their amplitude relatively small, so this value is very uncertain.

\subsection{Comparison with Predictions from Magnetospheric Models}

\indent We have investigated several magnetospheric models that appear in the literature and address the issue of differential rotation in magnetosphere and disk. Some models that approach the problem of magnetospheric accretion from the point of view of spin-up/spin-down of the young stars and jet formation are reviewed in Goodson \& Winglee (\cite{goodson} and references therein). They all consider the case of strong magnetic fields (i.e. there is disk truncation) and the physical distinction between them comes from the effective value of magnetic diffusion in the disk. In these models, the different rotation rates of the star and the disk cause the field lines that permeate the disk to become azimuthally twisted and a toroidal component of the magnetic field is created. The ability of the disk to counteract this toroidal component is described by the effective magnetic diffusivity.\\
\indent For the most diffusive flows, no wrapping up of the magnetic field lines occurs because the field lines can continuously slip through the disk. This is the case for the Ghosh \& Lamb (\cite{ghosh}) and the Shu et al. (\cite{shu}) models. According to Ghosh \& Lamb all the field lines are closed and they thread the disk up to distant radii. In the Shu et al. model the closed field lines thread only a small, nearly uniformly rotating, region of the disk and outside this region there are open magnetic field lines. \\
\indent For low magnetic diffusivity, differential rotation between the disk and the star makes the magnetic field lines wrap up and open via a mechanism known as inflation (Lovelace et al. \cite{lovelace}). The inner edge of the disk undergoes oscillations and magnetic reconnection occurs causing accretion episodes; no steady configuration is possible (Goodson \& Winglee \cite{goodson}).\\
\indent For intermediate diffusivity, once inflation opens the outer magnetic field lines, a steady state can be attained. The inner magnetic field lines remain closed and approximately in co-rotation with the star. The plasma in the disk can move radially inwards until it reaches the point where steady accretion occurs (Lovelace et al. \cite{lovelace}). In this last scenario, it is plausible that the closed magnetic field lines are azimuthally deformed.\\
\indent The time-lagged behaviour that we observe in SU~Aur suggests a
quasi-steady configuration, in which the magnetic field lines are
azimuthally curved as a result of the magnetic field line anchors being
dragged by, but slipping through the inner disk that is rotating somewhat
slower than the magnetic field line anchors on the stellar surface. The misalignment of rotation and magnetic axes introduces, at certain phases, an extra magnetic pressure component in the plane of the disk, that eases the loading of the closed magnetic field lines with matter from outside the co-rotation radius.\\
\indent But how do the outflows relate with this proposed scenario? Our proposed scenario includes open field lines that can drive wind flows (as suggested by Shu et al. \cite{shu}). We have found that the signatures of accretion and disk winds are about 240$\degr$ out of phase. This agrees with the suggestion of Johns \& Basri (\cite{johnsb}) (confirmed by Petrov et al. \cite{petrov}) if their picture of the oblique rotator is refined to include azimuthally curved magnetic field lines.


\section{Transient Blueshifted Absorption Features}

\subsection{Velocity Variations}

\begin{figure}[ht]
\resizebox{\hsize}{!}{\includegraphics{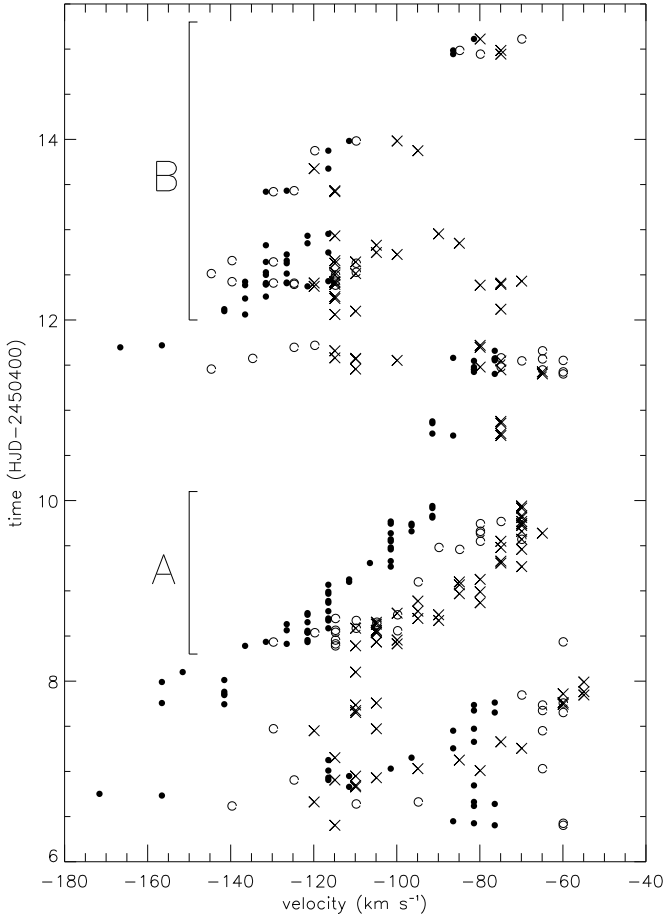}}
\caption{Velocity evolution of two drifting absorption features in the line profiles of $\bullet$ H$\alpha$, $\circ$ H$\beta$\ and $\times$ \ion{Na}{i}~D1. For \ion{Na}{i}~D2 the velocity position of the feature is similar to \ion{Na}{i}~D1. Feature~B was not detected in \ion{Na}{i}~D. Feature~A lasts for at least 1.6~days and feature~B for 3~days. The velocity position of the features is determined by detecting the local minimum in the line profiles. Simultaneously with feature~A, an intensity enhancement was observed in H$\alpha$\ and H$\beta$, affecting the whole profiles.}
\label{f6}
\end{figure}

\indent Two transient absorption features were detected in the blue wings of the profiles of H$\alpha$, H$\beta$\ and \ion{Na}{i}~D (Unruh et al. \cite{unruh98}). We did not pursue a multi-component analysis of these lines (as done in Giampapa et al. \cite{giampapa}, Johns \& Basri \cite{johnsb} or Oliveira \& Foing \cite{oliveiraa}) due to the optical thickness of the emission and the complexity and dynamics of the profiles. We could however isolate the transient features in the spectra by applying an adequate filtering procedure. The velocity position of the features is determined by detecting the minimum (in a chosen velocity interval) in the difference spectra with respect to the median-filtered spectra. In Fig.~\ref{f6}, the velocity evolution of these features is plotted (features~A and B). There is no considerable change in the width/depth of the features, but these measurements are very difficult to carry out due to the intrinsic variability of the underlying profile.\\
\indent Feature~A persisted for about 1.6~days. It drifted from approximately $-$140 to $-$90 km~s$^{-1}$\ in H$\alpha$\ and from $-$110 to $-$70 km~s$^{-1}$\ in \ion{Na}{i}~D1. In H$\beta$\ we cannot identify the velocity position of the feature unambiguously and so the minimum detection was more uncertain. However, there are indications that $|$v(H$\alpha$)$|$~$>$~$|$v(H$\beta$)$|$~$>$~$|$v(\ion{Na}{i})$|$. The velocity evolution in the three lines follows parallel tracks in time. The onset of the feature~A is simultaneous with the flux enhancement observed in H$\alpha$ and H$\beta$\ (Sect.\ 3.2).\\ 
\indent Approximately 4~days after the onset of feature~A, another transient feature was detected in H$\alpha$\ and H$\beta$\ but no counterpart was seen in the \ion{Na}{i}~D lines (feature B in Fig.~\ref{f6}). Feature~B may therefore originate in regions higher in the magnetosphere that contribute little to the \ion{Na}{i}~D line formation. In H$\alpha$\ this feature drifted from $-$140 to $-$80~km~s$^{-1}$ and seems to be decaying after 3 days when our time series ends. In contrast to what we observe for feature~A, the velocity position of feature~B in H$\beta$\ is practically the same as in H$\alpha$. We stress that feature~B is more conspicuous in H$\beta$. In the case of feature~B it is rather difficult to be certain if it is an absorption or emission transient, or a combination of both.

\subsection{Interpretation: Sporadic Mass Ejections}

\indent We propose that the first transient absorption feature might be plasma ejected upward, that condenses and afterwards decelerates under the influence of gravity, causing the observed velocity evolution. We have investigated several scenarios that could cause such velocity evolution. A blob of matter moving at constant velocity, in which case the velocity variation would be just the result of angular projection, is ruled out because it could not cause the velocity difference between the lines. The same problem is faced when postulating a blob that departs from the stellar surface with the escape velocity and then decelerates. The observed offset between the initial velocities of the lines indicates radially elongated material that underwent an acceleration outwards by an as yet unidentified mechanism, before it started to decelerate. For the second transient feature the velocity evolution shows no evidence of any initial acceleration. Still, as for the first feature, the velocity evolution strongly suggests deceleration even though the dispersion is very large.\\
\indent We investigated several types of solar-like phenomena that could have similar characteristics. Post-flare loops or surges are out of consideration because, although the matter is initially accelerated and condenses, it then falls back to the solar surface and would thus appear as redshifted absorption features, not blueshifted as we observe. Such redshifted absorption components have been detected on other stars though, e.g. \object{HR~1099} (Foing et al. \cite{foinga}) and \object{AD~Leo} (Houdebine et al. \cite{houdebine}) and were proposed to be the analogue of post-flare loops.\\
\indent If the deceleration measured for feature~A is interpreted as gravitational deceleration of material, then a fit to that deceleration rate indicates a location of the decelerating matter at approximately 7 to 8~R$_{\star}$. It is difficult to explain a column of material progressing at such distances from the star, specially without knowledge of its previous evolution.\\
\indent Feature~A might be the line-of-sight absorption of material of a large scale coronal mass ejection, in a configuration sometimes called a halo mass ejection. Similar to what is believed to happen in the Sun and flare stars (Foing \cite{foing}; Houdebine et al. \cite{houdebine}), a disruption of the magnetic field lines, perhaps near the truncation radius, might cause a ``forward'' mass ejection, away from the star, that cools and condenses and is observed as transient absorption features, and a ``backwards'' release of matter/energy, that can hit the star and cause the Balmer brightening we observe. In the case of SU~Aur, these ejections clearly point at a magnetospheric connection between the star and the disk where magnetic stresses and energy build up and are released sporadically. To reproduce the time evolution of the flare, as seen in the Balmer brightening, (peak at day~9.1, Fig.~\ref{f3}) and of the decelerated absorption component (start at day~8, Fig.~\ref{f6}), we suggest that the primary instability and energy release did not occur at the stellar surface or in the lower atmosphere, but rather at high altitude in the interacting magnetosphere between the star and the disk. For the second absorption feature, a similar scenario would apply, but without a visible detection of a flare.\\  
\indent The events that we observed are very likely triggered by magnetic field disruption, perhaps in the context of sheared magnetic field lines as described in the previous section. The energy released during these events, in combination with more continuous energy release from magnetic field reconfiguration, might explain the especially high X-ray luminosity of SU~Aur (Skinner \& Walter \cite{skinner}).


\section{Summary}

\indent The circumstellar environments of CTTS are extremely interesting for studying the interaction between the star and the disk. The CTTS line profiles are very complex and diverse, as both accretion and outflows play a role in their formation. In this paper we try to disentangle to some extent these effects, by analysing H$\alpha$, H$\beta$, \ion{Na}{i}~D and \ion{He}{i}~D3, in an attempt to constrain the geometry and dynamics of the magnetosphere.\\
\indent The \ion{He}{i}~D3 line in SU~Aur has two components. Their variability is slightly time-lagged, indicating spatially displaced regions of formation. The component at rest velocity can be formed close to or at the stellar surface, for instance in the post-shock region, and the redshifted component in the accretion column. We propose that the redshifted component of \ion{He}{i}~D3, the red wings of \ion{Na}{i}~D and the redshifted absorption features in H$\beta$\ and H$\alpha$\ are all formed in the accretion funnel, at different distances from the star according to their opacities and excitation conditions. We find different time lags between these lines, with the variability of the lines closer to the star preceding the variability in the outer spectral lines. We argue that such behaviour cannot be caused by non-steady accretion, nor by simple occultation effects in an oblique magnetospheric model. We propose that such time-lagged behaviour can only occur if the magnetic field lines in the oblique magnetosphere have an azimuthal component.\\
\indent We find that the near blue wings of H$\alpha$, H$\beta$\ and  \ion{Na}{i}~D (wind indicators) are out of phase with the accretion indicators, though not exactly in anti-phase.\\
\indent We observe two transient absorption features in the blue wings of H$\alpha$, H$\beta$\ and \ion{Na}{i}~D. The velocity evolution of the first feature indicates matter to be initially accelerated (different velocities for the different lines). Subsequent evolution suggests a deceleration or a projection effect. We propose that these features are the signature of plasma outflow events, powered by the disruption of the distorted magnetic field lines.

\begin{acknowledgements}
JMO research work is being supported by the Praxis XXI grant BD9577/96 from the {\it Funda\c{c}\~{a}o para a Ci\^{e}ncia e a Tecnologia}, Portugal. YCU a\-cknow\-led\-ges the support through grant S7302 of the Austrian {\it Fond zur Wissenschaftlichen F\"{o}rderung}. The authors acknowledge the staff of the observatories involved and the other members of the MUSICOS 96 collaboration (in alphabetical order): T.~B\"{o}hm, H.~Cao, C.~Catala, A.~Collier Cameron, J.F.~Donati, P.~Ehrenfreund, J.~Hao, A.P.~Hatzes, H.F.~Henrichs, L.~Huang, C.M.~Johns-Krull, J.A. de Jong, L.~Kaper, E.J.~Kennelly, E.~ten Kulve, J.~Landstreet, N.~Morrisson, C.L.~Mulliss, J.E.~Neff, R.S.~Le~Poole, C.~Schrijvers, T.~Simon, H.C.~Stempels, J.H.~Telting, N.~Walton and D.~Yang. We thank the referee G.~Basri for valuable comments.

\end{acknowledgements}

\appendix

\section{Cross-Correlation Function Analysis}

\subsection{Definition and Computational Method}

\indent Let us assume two stationary time series L$_{1}$(t) and L$_{2}$(t) discretely sampled N times at times t$_{i}$, with  $\Delta$t~=~t$_{i+1}-$t$_{i}$ for all values $1\le i\le$ N$-$1. The {\it cross-correlation function } ($CCF$), computed at intervals $\tau$ (or time lags) which are integer multiples of the sampling interval $\Delta$t, is defined as follows:

\begin{equation} CCF(\tau)~=~\frac{1}{\mathrm{N}} \sum_{i~=~1}^{\mathrm{N}-1}
\frac{[\mathrm{L}_{1}(\mathrm{t}_{i})-{\overline \mathrm{L}}_{1}] [\mathrm{L}_{2}(\mathrm{t}_{i}-\tau)-{\overline
\mathrm{L}_{2}}]}{\sigma_{1} \sigma_{2}}
\end{equation}

\noindent where ${\overline \mathrm{L}}_{\mathrm{k}}$ and $\sigma_{\mathrm{k}}$ are respectively the mean and standard deviation of each time series L$_{\mathrm{k}}(\mathrm{t}_{i})$. The obvious problem is that our time series is unevenly spaced, so this definition has to be adapted.\\
\indent There are two methods to circumvent this difficulty: the {\it interpolation method} and the {\it discrete correlation function}. In the first method (Gaskell \& Peterson \cite{gaskell}), the $CCF$ is calculated twice for each time lag $\tau$: the observed points L$_{1}$(t$_{i})$ are correlated with the interpolated values of L$_{2}$(t$_{i}-\tau)$ and the observed points L$_{2}$(t$_{i})$ are correlated with the interpolated values L$_{1}$(t$_{i}+\tau)$. The two results are then averaged, assuming there is no reason to prefer one interpolation over the other. The second method, the discrete correlation function ($DCF$) was introduced by Edelson \& Krolik (\cite{edelson}). In this method the cross-correlation function is computed at each time lag using only real data points separated by $\tau$. The resulting correlation function is then binned on intervals $\delta$t so that the calculated value at time lag $\tau$ is an average over the interval $\tau~\pm~\delta$t/2.\\
\indent These methods were found to be in good agreement for well sampled data sets with a large number of data points (White \& Peterson \cite{white}, henceforth WP, and references therein). The $DCF$ method is more general in the sense that it makes no assumptions on the nature of the time series that we are sampling. When using the interpolation method, it is assumed implicitly that the sampled time series vary smoothly on the time scales that correspond to the intervals between observations. Still, the $DCF$ will require more points than the interpolation method to give a meaningful result and, in the limit of poor sampling, will more easily fail to give a significant lag. Therefore, for our data set we decided that the {\it interpolation method} is more adequate and hence used this in the calculations presented in this paper.

\subsection{Optimization Techniques}

\indent When calculating the $CCF$ for several types of test functions with the same time sampling as our data points, we realised that the calculations yielded $CCF$ peak values above unity. There was clearly a normalization problem in the calculation of the $CCF$. For stationary time series, the mean and variance of the sample are invariant in time; only in this situation it is valid to calculate ${\overline \mathrm{L}}_{\mathrm{k}}$ and $\sigma_{\mathrm{k}}$ using the entire time series. We have concluded that this is not always a reasonable assumption: in fact, in our data set there was a strong flux enhancement in the spectral lines (Sect.\ 3.2), so clearly these statistical quantities change in time. Thus, we concluded that it was more appropriate to compute ${\overline \mathrm{L}}_{\mathrm{k}}$ and $\sigma_{\mathrm{k}}$ for each lag $\tau$, with only the data points effectively used for the computation of $CCF$($\tau$). Therefore, the calculation of $CCF$($\tau$) reduces to the calculation of the linear correlation coefficient at each lag $\tau$.\\
\indent We address now the question of how the interpolation is actually performed. First, following WP, we do not interpolate beyond the limits of the time series: we exclude from the calculation of $CCF$($\tau$) points where t$_{i} - \tau <$ t$_{1}$ or t$_{i} - \tau >$ t$_{\mathrm{N}}$. We also use linear interpolation of the data points (WP). As it is mentioned above, the interpolation assumes that the times series vary smoothly between the sampled data points. In our time series the extreme data points are separated from the rest of the observations by more that 1.5~days. We have computed the $CCF$ both including and excluding these points and we concluded that the $CCF$ peaks are much sharper in the last case. So we have excluded those data points from $CCF$ calculation, even though we still use them for the calculation of the interpolated series. We should point out that the number of points used is only decreased to N$-$4.\\
\indent As we have mentioned the sampling of H$\beta$\ is worse: we have only 72 H$\beta$\ spectra against 126 for the remaining lines. The calculated $CCF$($\tau$) is an average quantity if there is no reason to favour one of the interpolations. But if we are interpolating H$\beta$\ and H$\alpha$, for example, it is clear that the interpolation of the H$\alpha$\ series is better because it is computed with more data points. Thus, when a $CCF$ that involves H$\beta$\ is computed we will consider only one term, the one where the other time series is interpolated. In the example above we consider only the terms where the interpolated H$\alpha$\ series is cross-correlated with the observed H$\beta$\ series.

\subsection{Error Analysis}

\indent Gaskell \& Peterson (\cite{gaskell}) propose an analytical formula to compute the uncertainty of the time lag $\tau$. WP and  Maoz \& Netzer (\cite{maoz}) found that this formula can severely underestimate the uncertainties. Accordingly, we have decided to use Monte Carlo simulations to estimate these errors. We assume that the noise in our time series has a gaussian distribution. We use the standard deviation of the continuum as a conservative estimate of the noise in the spectral lines. We computed the $CCF$ with a sampling step of $\Delta \tau=0.025$~day (approximately half of the maximum exposure time in the time series). We used 1000 trials for the simulations. We then computed the {\it empirical cross-correlation peak distribution} ($CCPD$) (Maoz \& Netzer \cite{maoz}), by simply counting how many times the peak occurred at each time lag. The peak of the $CCPD$ is the most likely time lag and its width is the uncertainty of this estimate. Given the asymmetry of the $CCPD$, we follow WP and define the quantity $ \pm \Delta \tau _{67}$, for which 2/3 of the trials give a lag in the interval $\tau \pm \Delta \tau _{67}$ with relation to the median of the distribution. When the $CCPD$ is a gaussian $\Delta \tau _{67}$ corresponds to the standard deviation. The quantity $\Delta \tau _{67}$ is derived directly from the cumulative $CCPD$.

\end{document}